\documentclass[12pt]{article} 
\usepackage{graphicx,subfigure,latexsym,amssymb}
\usepackage{bm}

\newcommand{\be}{\begin{equation}}
\newcommand{\ee}{\end{equation}}
\newcommand{\bear}{\begin{eqnarray}}
\newcommand{\eear}{\end{eqnarray}}
\newcommand{\ba}{\begin{array}}
\newcommand{\ea}{\end{array}}

\def\({\left(}
\def\){\right)}

\begin{document}

\begin{titlepage}
\vfill
\begin{flushright}
{\normalsize RBRC-1245}\\
\end{flushright}

\vfill
\begin{center}
{\Large\bf Shear Viscosity of Quark-Gluon Plasma in Weak Magnetic Field in Perturbative QCD: Leading Log }

\vskip 0.3in
Shiyong Li$^{1}$\footnote{e-mail: {\tt  sli72@uic.edu}} and
Ho-Ung Yee$^{1,2}$\footnote{e-mail:
{\tt hyee@uic.edu}}
\vskip 0.3in

 {\it $^{1}$Department of Physics, University of Illinois,} \\
{\it Chicago, Illinois 60607 }\\[0.15in]
{\it $^{2}$RIKEN-BNL Research Center, Brookhaven National Laboratory,} \\
{\it Upton, New York 11973-5000 }
\\[0.3in]

{\normalsize  2017}

\end{center}

\vfill

\begin{abstract}

We compute the shear viscosity of two-flavor QCD plasma in an external magnetic field in perturbative QCD at leading log order, assuming that the magnetic field is weak or soft: $eB\sim g^4\log(1/g)T^2$.
We work in the assumption that the magnetic field is homogeneous and static, and the electrodynamics is non-dynamical in a formal limit $e\to 0$ while $eB$ is kept fixed.
We show that the shear viscosity takes a form $\eta=\bar\eta(\bar B)T^3/(g^4\log(1/g))$ with a dimensionless function $\bar\eta(\bar B)$ in terms of a dimensionless variable $\bar B=(eB)/(g^4\log(1/g)T^2)$. The variable $\bar B$ corresponds to the relative strength of the effect of cyclotron motions compared to the QCD collisions: $\bar B\sim l_{mfp}/l_{cyclo}$. We provide a full numerical result for the scaled shear viscosity $\bar\eta(\bar B)$.

\end{abstract}

\vfill

\end{titlepage}
\setcounter{footnote}{0}

\baselineskip 18pt \pagebreak
\renewcommand{\thepage}{\arabic{page}}
\pagebreak

\section{Introduction \label{sec1}}

The transport properties of nuclear matter under extreme temperature or density that are ultimately described by Quantum Chromo Dynamics (QCD) play important roles in our understanding of dynamics of the Quark-Gluon Plasma (QGP) created in relativistic heavy-ion collision experiments and dense quark-matters that could possibly be formed at the center of neutron stars. The relevant space-time scales in real-time descriptions of these systems (or the inverse scales of momentum and energy of inhomogeneity) are reasonably assumed or argued to be much larger than the scale of the microscopic QCD interactions that may be called an effective mean-free path. In this case, the powerful universal framework of hydrodynamics, that needs only a small number of inputs such as equation of states and a few transport coefficients together with the conservation laws of energy-momentum and charges, can be used to 
describe the interesting real-time evolution of the system (see Refs.\cite{Kolb:2003dz,Romatschke:2009im,Jeon:2015dfa} for recent reviews). Most of the current realistic numerical simulations of the QGP in heavy-ion collisions and the nuclear matter of (merging) neutron stars are based on the hydrodynamics framework, at least in the major parts. 

One of the most important transport coefficients in these hydrodynamics descriptions is the shear viscosity that governs the rate of momentum transfer in a presence of inhomogeneity of fluid velocity, while the bulk viscosity describes the change of local pressure when the fluid element is either expanding or contracting. In the case of QGP in relativistic heavy-ion collisions, a detailed comparison between the hydrodynamic predictions and the experimental data on the momentum anisotropy of emitted particles (called elliptic flow) in the azimuthal angle perpendicular to the beam direction \cite{Teaney:2000cw,Schenke:2012wb} indicates that the shear viscosity to entropy density ratio of QGP is close to probably the smallest value that has ever been realized in Nature \cite{Kovtun:2004de}; $\eta/s\sim 1/4\pi$, that could only be possible when the QCD is strongly coupled in these systems, that is, the coupling constant of QCD is not small.

Although the theoretical and experimental analysis of these systems to date strongly suggests that the dynamics of QCD in the relevant regimes of these systems is strongly coupled in most history of the time evolution, the QCD in asymptotically high temperature is weakly coupled and the perturbative theory in finite temperature becomes valid. Although the values of transport coefficients in this perturbative regime are of little significance to the actual experiments, the computation of transport coefficients in this high temperature perturbative regime \cite{Baym:1990uj,Heiselberg:1994px,Arnold:2000dr,Arnold:2003zc} have given us useful insights on the underlying physics of transport coefficients, and could also provide useful bench-mark information on the value of the transport coefficients in one extreme end, compared to those in the other end of infinitely strong coupling that may be computed in the AdS/CFT correspondence \cite{Kovtun:2004de}. On a more practical side, the perturbative regime is where one can perform a systematic and reliable theoretical computation typically allowing even analytic results, at least up to certain lowest orders.

In this work, we compute the shear viscosity of QCD plasma in the presence of magnetic field in high enough temperature, so that perturbative QCD can apply. The QCD plasma or nuclear matter in magnetic field has been studied extensively recently, due to its relevance in off-central heavy-ion collision experiments and in rotating compact stars (magnetars) (see Refs.\cite{Hattori:2016emy,Miransky:2015ava,Kharzeev:2013jha} for the recent reviews). Our proposed study of QGP in magnetic field in high enough temperature is motivated by the off-central heavy-ion collisions, where a huge magnetic field, albeit transient, is present and may affect the properties of the created QGP fireball \cite{Kharzeev:2007jp}. The typical strength of the magnetic field at very early time is $eB\sim (200-400\,\, MeV)^2$ \cite{Skokov:2009qp,Deng:2012pc,Bzdak:2011yy,Bloczynski:2012en}, while it becomes smaller by a factor 10 roughly after 1 fm/c, beyond which it remains more or less constant over a few fm/c due to the Lenz effect \cite{Tuchin:2014iua,McLerran:2013hla}. Compared to the expected temperature of the QGP of about $250-400$ MeV, the scale of magnetic field is comparable to the temperature initially, but becomes smaller in most of the stages after 1 fm/c. This motivates our assumption that the strength of magnetic field is parametrically smaller than the temperature (that is, $eB\ll T^2$) in powers of coupling constant $g\ll 1$ in our perturbative QCD computation.
We should point out that our assumption with a particular $g$-dependence of magnetic field that is described  in section \ref{sec2} is not entirely motivated by the experiments, but also by the practical reason that we can perform a reliable and interesting (to our eyes, at least) perturbative QCD computation with this assumption.
Up to our knowledge, there has not been a systematic perturbative QCD computation of shear viscosity either in weak magnetic field limit that we study in this work or in strong magnetic field limit (that is, $eB
\gg T^2$). For the recent computation of bulk viscosity in strong magnetic field limit, see Ref.\cite{Hattori:2017qih}. See also
Refs.\cite{Fukushima:2015wck,Li:2016bbh,Hattori:2016cnt,Hattori:2016lqx,Fukushima:2017lvb} for the computations of other transport coefficients in strong or moderate magnetic field limit in perturbative QCD, and Ref.\cite{Finazzo:2016mhm} for the shear viscosity in magnetic field in the AdS/CFT correspondence for strong coupling regime.

We also consider a simplified set-up where the magnetic field is homogeneous and static. This also means that we neglect back reactions from the plasma to the electromagnetic fields, that is, we do not include dynamics of electrodynamics, and the magnetic field is treated as an external environment (for example, there is no induced electric field). This approximation can be formally justified in the limit where the electric coupling that governs the back reaction goes to zero $e\to 0$ while $eB$ that governs the effect of the magnetic field to the plasma remains fixed. Note, however that the Lenz effect mentioned in the above is due to the back reaction from the plasma to the magnetic field. Therefore, this limit we consider is 
purely of theoretical nature, not driven by any experimental conditions. 
We will show in the next section that in this case, the components of fluid velocity transverse to the magnetic field direction {\it freeze} to zero, that is, they decay with a finite relaxation rate even in a zero spatial gradient limit, so that they are no longer long-lived hydrodynamics variables in the emerging new hydrodynamics at a sufficient low momentum regime. 
We will identify the transport coefficients in this new effective hydrodynamics at low energy, and show that there is only one shear viscosity (and two bulk viscosities) surviving in this regime. We compute this shear viscosity in this work.

If one considers a more realistic case of dynamical electromagnetism coupled to the plasma, it is well known that there emerges a new hydrodynamics at scales below $k<\sigma$ ($\sigma$ is the conductivity), coined as magnetohydrodynamics (MHD) (see Refs.\cite{Huang:2009ue,Huang:2011dc,Hernandez:2017mch,Grozdanov:2016tdf,Hattori:2017usa} for recent developments of relativistic MHD). The hydrodynamics variables of MHD are the magnetic field in local rest frame, $B^\mu\equiv \epsilon^{\mu\nu\alpha\beta}u_\nu F_{\alpha\beta}$, and the fluid velocity $u^\mu$, while the electric field in the local rest frame $E^\mu=u_\nu F^{\mu\nu}$ and the local charge density  $n=u_\mu j^\mu$ decay to zero with a finite relaxation time $\tau\sim 1/\sigma$ and are thus excluded in the hydrodynamics variables of MHD by the same reason as above. Note that in the lab frame, $E^\mu\propto \vec E+\vec v\times\vec B$. In more physical terms, what is happening is that for any $\vec v$ and $\vec B$, the plasma back-reacts to Lorentz force via induced currents to ensure that $\vec E$ relaxes to the {\it local equilibrium value} $\vec E_{eq}=-\vec v\times\vec B$ within a time scale $1/\sigma$ (it is the local equilibrium value, since $E^\mu$ is the electric field in the local rest frame of fluid which should vanish in order to be in equilibrium). 
In the presence of a background magnetic field $\vec B_0$ in this set-up, this essentially fixes the electric field fluctuations in terms of transverse velocity fluctuations to the magnetic field at linearized level: $\delta \vec E=-\delta\vec v\times\vec B_0$. The ensuing MHD equations of motion give rise to a propagating wave of these fluctuations along the background magnetic field, called Alfven wave with velocity $v_A^2=B_0^2/(\epsilon+p+B_0^2)$ ($\epsilon$ and $p$ are energy density and pressure).
We should also point out that there are currently much efforts to simulate the heavy-ion collisions with dynamical electromagnetism coupled to the plasma fireball \cite{Inghirami:2016iru,Roy:2017yvg}, aiming at reliable theoretical predictions of various transport phenomena associated to magnetic and electric fields, such as the Chiral Magnetic Effect \cite{Fukushima:2008xe,Huang:2015fqj}, the charge dependent elliptic flows \cite{Burnier:2011bf,Gorbar:2011ya} and the slope of $v_1$ in rapidity due to Lorentz force \cite{Gursoy:2014aka}.

In our simplified limit of non-dynamical electromagnetism ($e\to 0$), we don't have such MHD regime, noting that  $\sigma$ is proportional to $e^2$ in our convention, so the time scale for the MHD is arbitrary long and is never realized.
Instead, we can make the following connection between the above discussion and our limit of non-dynamical electromagnetism: in our limit of non-dynamical electromagnetism, let's work in the lab frame where $\vec E=0$ always, while $\vec B=\vec B_0$ is a fixed external field.
Making $E^\mu\sim \vec v\times\vec B_0=0$ for local equilibrium is only possible with the transverse components of velocity to the magnetic field vanishing. This is the ultimate reason why the transverse velocity is excluded in the low energy description in our limit (note that the transverse velocity is {\it not} excluded in the MHD. Only the combination $\vec E+\vec v\times\vec B_0$ is excluded in the MHD and $\vec v$ can be an independent MHD variable). However, the relaxation mechanism (and therefore the relaxation time) is different from that in the MHD via dynamical Maxwell's equation: in our case, it will be shown in Eq.(\ref{rex}) that it comes solely from the induced current and the associated Lorentz force (without any dynamical Maxwell equations) with the relaxation time $\tau\sim {(\epsilon+p)\over\sigma B_0^2}$, which is finite in our limit recalling that $\sigma$ contains $e^2$ and $eB_0$ is finite in our $e\to 0$ limit.
The new low energy hydrodynamics in our limit appears in the scales $k<1/\tau$. 

In summary, in the case of dynamical electromagnetism (with or without a background magnetic field), there appears a new low energy hydrodynamics, called the MHD in the scales $k<1/\sigma$. The hydrodynamical variables of the MHD are significantly reduced compared to the original ``microscopic hydrodynamics'' that is valid when $k>1/\sigma$.
In our case of non-dynamical electrodynamics ($e\to 0$) with a finite background magnetic field $eB_0\neq 0$,
there appears a new low energy hydrodynamics at the scales $k<{\sigma B_0^2\over\epsilon+p}$.
The hydrodynamics variables in this low energy hydrodynamics are also reduced compared to the microscopic hydrodynamics, and one such reduction is the absence of the transverse velocity to the background magnetic field (if there is no other external electromagnetic field applied). As explained in the paragraph around Eq.(\ref{cur}) in the next section, another reduction is the absence of transverse charge current and hence the transverse conductivity in a neutral plasma\footnote{If the plasma has a non-zero charge density $n$, there exists a Hall current $\vec j=n\vec v_{eq}=n{\vec E_\perp\times\vec B_0\over B_0^2}$ in response to an external small electric field $\vec E_\perp$. See for example, Ref.\cite{Fukushima:2017lvb} for a recent study. We focus on a relativistic neutral plasma in this work where such Hall effect is absent.}.

We should emphasize that our consideration of hydrodynamics in a background magnetic field is not at all new, and there are several previous studies on the subject with notable progress \cite{Huang:2009ue,Huang:2011dc,Hernandez:2017mch,Grozdanov:2016tdf}, and it is pertinent to summarize how our low energy effective hydrodynamics and the corresponding shear viscosity is related to those studies.
One important criterion in this comparison is whether electrodynamics is assumed to be dynamical or non-dynamical. For example, Ref.\cite{Finazzo:2016mhm} works with non-dynamical electrodynamics, the Refs.\cite{Huang:2009ue,Huang:2011dc} consider the dynamical case, and the Ref.\cite{Hernandez:2017mch} considered both non-dynamical and dynamical cases. In the non-dynamical case, our result for the relaxation time $\tau\sim {(\epsilon+p)\over\sigma B_0^2}$ in fact agrees the one in Ref.\cite{Hernandez:2017mch}, for example Eq.(3.14) in a neutral case. The difference is not about the result, but instead about what energy regime we are focusing on, and what effective description we have for different regimes. Since the dispersion Eq.(3.14) in Ref.\cite{Hernandez:2017mch} is not truly hydrodynamic as it doesn't vanish in $k\to 0$ limit, this mode corresponding to transverse velocity is excluded in a sufficiently low momentum regime. In the dynamical case, on the other hand, since the full components of fluid velocity remain as the hydrodynamics variables in the MHD, and a background magnetic field breaks rotational invariance, there naturally appear five different shear viscosities corresponding to five different shear gradient modes with respect to the magnetic field direction, nicely classified in Refs.\cite{Huang:2009ue,Huang:2011dc,Hernandez:2017mch}.
As explained in the above, the MHD regime is pushed to zero momentum (that is, $k<\sigma\to 0$) in the limit of non-dynamical electromagnetism ($e\to 0$), and there is no overlap between the MHD regime and the regime of our low energy effective hydrodynamics in the non-dynamical limit: $k<{\sigma B_0^2\over (\epsilon+p)}$, so there is no logical correspondence that can be made between these five shear viscosities and our shear viscosity.
Computation of these five shear viscosities in the MHD regime therefore remains as an open problem. 

However, in either case, if the momentum scale is higher than the relaxation scale of $\sigma$ or ${\sigma B_0^2\over (\epsilon+p)}$, so that we go to the regime of ``microscopic" hydrodynamics, the full component of velocity is included in the hydrodynamics variables, and the five different shear viscosities make sense in this ``UV" regime. The results in Refs.\cite{Finazzo:2016mhm,Hernandez:2017mch} for the non-dynamical case should be viewed in this way.
Then, our shear viscosity can be considered as the low energy limit of (one of) those five ``UV'' shear viscosities,
while the other four viscosities lose their meaning in the new low energy hydrodynamics (in fact, two of the five viscosities in Ref.\cite{Huang:2011dc}, $\eta_3$ and $\eta_4$, are odd under charge-conjugation and they vanish identically in the neutral plasma $n=0$ that we are considering, irrespective of the regimes. If we relax the neutrality condition, we also show in the Appendix that we have one more shear viscosity coming from $\eta_4$ of Ref.\cite{Huang:2011dc} surviving in our low energy regime. The same is true for the Hall conductivity, see the footnote in the previous page).  

In summary, the five shear and two bulk viscosities in the Refs.\cite{Finazzo:2016mhm,Hernandez:2017mch} in the case of non-dynamical electromagnetism are for the ``microscopic hydrodynamics" in 
high momentum regime, and the framework needs to be replaced by our new hydrodynamics at sufficiently low momentum scales. In the Appendix, we show explicitly how the five shear and two bulk viscosities nicely classified in Ref.\cite{Huang:2011dc} reduce to our one shear and two bulk viscosities in a {\it neutral} plasma when we remove the transverse components of fluid velocity to the magnetic field direction in our low energy limit (Note that the classification in Ref.\cite{Huang:2011dc} relying only on the tensor structures and symmetries can be applied to any regime, either in the MHD regime or in the ``UV'' regime).

\section{Emergent low energy hydrodynamics and shear viscosity in an external magnetic field}

In this section, we show that in a presence of an external non-dynamical magnetic field, there emerges a new effective low energy hydrodynamics at sufficiently low scales and the hydrodynamical variables are reduced in this low energy hydrodynamics. Specifically, we will argue that the transverse components of velocity perpendicular to the magnetic field disappear in this regime, and there is only one meaningful notion of shear viscosity in this effective hydrodynamics emerging in sufficiently low energy regime.
 
As in the conventional approach, hydrodynamics describes the modes that live arbitrary long when their spatial gradient is arbitrary small. This can only be possible if these modes correspond to the parameters characterizing possible equilibrium states, so that they stay constant when they are homogeneous.
All other modes have finite relaxation times even in the homogeneous limit, and are excluded in the hydrodynamic description in sufficiently low energy regime: they are quasi-normal modes.
Therefore, we should first identify all equilibrium parameters of the plasma in the presence of an external magnetic field, in order to construct a hydrodynamics in a non-dynamical magnetic field. Let the magnetic field point to $z$-direction in a fixed lab frame, $e{\bm B}=eB\hat{\bm z}$.
Detailed balance with energy conservation gives us the temperature $T$ as one of such parameters.
Without a magnetic field, the momentum conservation would give us a vector parameter, the fluid velocity $\bm v$, as another parameter for equilibrium. However, in the presence of an external magnetic field, the transverse momentum perpendicular to $\hat z$ direction is not conserved due to Lorentz force, and only the longitudinal momentum along the magnetic field direction is conserved. We therefore expect that the longitudinal fluid velocity, $v_\parallel=v_z$, remains as an equilibrium parameter, but ${\bm v}_\perp$ is no longer an equilibrium parameter, and should be excluded in the emerging low-energy hydrodynamics with an external magnetic field. We show that ${\bm v}_\perp$ indeed becomes a quasi-normal mode in the following.
 
Let us assume the existence of conventional hydrodynamics valid in the regime where the magnetic field can be treated as a first order in gradient expansion. We call this ``microscopic hydrodynamics". The 4-velocity $u^\mu=(\gamma,\gamma{\bm v})$ is a hydrodynamic variable in this microscopic hydrodynamics.
The current in this regime is given by 
\be
j^\mu=\sigma E^\mu\,,\quad  E^\mu\equiv F^{\mu\nu}u_\nu\,,
\ee
where $\sigma$ is the conductivity. In non-relativistic limit $\gamma\approx 1$ with a finite ${\bm v}_\perp$,
the spatial component of the current becomes,
\be
\bm j=\sigma \bm v_\perp\times \bm B\,,
\ee
whose origin is nothing but the Lorentz force in the lab frame, or equivalently the Ohmic current in the rest frame of the fluid.
From the transverse component of the energy-momentum conservation, $\partial_\mu T^{\mu\nu}=F^{\nu\alpha}j_\alpha$ (which originates from Lorentz force again), and the constitutive relation $T^{0\perp}=(\epsilon+p)\bm v_\perp$,
we have
\be 
 \partial_t \bm v_\perp={1\over (\epsilon+p)} \bm j\times {\bm B}=-{\sigma B^2 \over (\epsilon+p)}\bm v_\perp\equiv -{1\over \tau_R}\bm v_\perp\,,\label{rex}
\ee
which means that $\bm v_\perp$ is a quasi-normal mode with a relaxation time $\tau_R$.
In the emerging low-energy hydrodynamics in the energy regime smaller than $1/\tau_R$, we no longer have ${\bm v}_\perp$ as a hydrodynamic variable.

We consider only neutral plasma in this work, and the low energy hydrodynamic variables are $T$ and $v_z$ (or equivalently 1+1 dimensional velocity vector $u^\mu_\parallel$ where $\mu$ runs only along $(t,z)$ and it is normalized as $u_\mu u^\mu=-1$), which vary slowly in space-time: $T(x)$, $u^\mu_\parallel(x)$. Note that the variation can happen along the transverse direction as well as in the longitudinal direction.
In the Appendix, we show that the energy-momentum tensor up to first order in gradient generally takes a form,
\be
T^{\mu\nu}=(\epsilon+p_\parallel)u^\mu_\parallel u^\nu_\parallel +p_\parallel g^{\mu\nu}_\parallel+p_\perp g^{\mu\nu}_\perp-\eta (\partial_\perp^\mu u^\nu_\parallel+\partial_\perp^\nu u^\mu_\parallel)-(\zeta (u^\mu_\parallel u^\nu_\parallel+g^{\mu\nu}_\parallel)+\zeta' g^{\mu\nu}_\perp)( \partial_{\parallel\alpha} u^\alpha_\parallel)\,,\label{consti}
\ee
with one shear viscosity $\eta$ and two bulk viscosities $(\zeta,\zeta')$. This $\eta$ is what we compute in this work. Our metric convention is $g^{\mu\nu}=(-1,+1,+1,+1)$ where the first two indices correspond to 1+1 dimension of $(t,z)$ and the last two the transverse dimensions $x_\perp=(x,y)$. We also define $g_\parallel^{\mu\nu}=(-1,1,0,0)$, $g_\perp^{\mu\nu}=(0,0,1,1)$, $\partial_\parallel^\mu=(-\partial_t,\partial_z,0,0)$, $\partial_\perp^\mu=(0,0,\partial_x,\partial_y)$, and $u^\mu_\parallel=\gamma(1,v_z,0,0)$ where $\gamma=(1-v_z^2)^{-{1\over 2}}$. The combination $(u^\mu_\parallel u^\nu_\parallel+g^{\mu\nu}_\parallel)\equiv\Delta^{\mu\nu}_\parallel$ is the rest-frame space projection operator orthogonal to $u^\mu_\parallel$ in 1+1 dimensions, and $g_\perp^{\mu\nu}=\Delta^{\mu\nu}_\perp$ is the space projection to the transverse $(x,y)$ dimensions (see Appendix for a detailed discussion on these).
In the same Appendix, we also make a detailed connection between the above viscosities and the previous five shear and two bulk viscosities introduced/classified in Ref.\cite{Huang:2011dc}.
 If we take only the first three ideal parts, we simply have in components in the rest frame of the fluid (where $u^\mu_\parallel=(1,0,0,0)$)
\be
T^{\mu\nu}_{ideal}=(\epsilon,p_\parallel,p_\perp,p_\perp)\,,
\ee
which is the most general form of the ideal energy-momentum in a rest frame with a magnetic field (recall that the magnetic field is invariant under the boost along its direction, so it is the same in an arbitrary fluid rest frame). 
The ideal part of (\ref{consti}) is obtained by boosting this along the $z$ direction by $u^\mu_\parallel$.
In non-relativistic limit, the shear viscosity appears as
\be
T^{\perp z}=-\eta \partial_\perp v_z\,,\label{shear}
\ee
and our computation in the following sections is based on this.

Although this is not a subject we study in this work, we can follow the same logic to find that there is no notion of transverse electric conductivity in the emerging low energy hydrodynamics with an external magnetic field.
Let's apply a small transverse electric field ${\bm E}_\perp$ so that $|\bm E_\perp|\ll |\bm B|$.
The microscopic hydrodynamics gives the current from $j^\mu=\sigma E^\mu$ as
\be 
\bm j=\sigma(\bm E_\perp+\bm v_\perp\times \bm B)\,,
\ee
and the energy-momentum conservation gives the equation
\be
\partial_t \bm v_\perp={1\over (\epsilon+p)} \bm j\times {\bm B}={\sigma\over\epsilon+p}\left({\bm E}_\perp\times\bm B-B^2 \bm v_\perp\right)= -{1\over \tau_R}(\bm v_{eq}-\bm v_\perp)\,,
\ee
where the equilibrium transverse velocity $\bm v_{eq}$ is
\be
\bm v_{eq}={\bm E_\perp\times\bm B\over B^2}\,.
\ee
The above equation tells us that the transverse velocity relaxes to $\bm v_{eq}$ with the same relaxation time $\tau_R$, that is, any deviation from $\bm v_{eq}$ is a quasi-normal mode in the low energy hydrodynamics. With $\bm v_{eq}$, the current vanishes
\be
\bm j=\sigma(\bm E_\perp+\bm v_{eq}\times \bm B)=0\,.\label{cur}
\ee
The physics behind this result is the following: the frame moving with velocity $\bm v_{eq}$ is precisely the frame where the electric field vanishes and there exists only a non-zero magnetic field. We know that in this frame, the transverse velocity should relax to zero with the relaxation time $\tau_R$ according to our previous discussion. In the original lab frame, this is equivalent to the relaxation of the velocity to $\bm v_{eq}$. Since there is no electric field in the equilibrium rest frame, the current vanishes in this frame and hence in the original frame as well (recall that our plasma is neutral).

The situation is different when $|\bm E_\perp|>|\bm B|$ and there is no such frame where electric field vanishes. Instead, we have a frame moving with velocity $\bm v_\perp={\bm E_\perp\times\bm B\over E_\perp^2}$ where the magnetic field vanishes, and the electric field becomes
\be
\bm E_\perp'=\bm E_\perp \left(1-{B^2\over E_\perp^2}\right)\,.
\ee
We then have a current
\be
\bm j=\sigma \bm E_\perp'=\sigma \bm E_\perp \left(1-{B^2\over E_\perp^2}\right)\Theta(E_\perp-B)\,.
\ee

\section{QCD Boltzmann equation in leading log \label{sec2}}

In this section, we describe the technical details of our computation of shear viscosity in weak/soft magnetic field in perturbative regime of QCD.
More explicitly,
we assume that the magnetic field is weak or soft in the sense that its scale is comparable to the effective 2-to-2 QCD collision rate in leading log, or more precisely,
\be
eB\sim g^4\log(1/g)T^2\ll T^2\,.
\ee
In this case, we will show that the shear viscosity in leading log takes a form
\be
\eta=\bar\eta(\bar B){T^3\over g^4\log(1/g)}\,,\quad \bar B\equiv {eB\over g^4\log(1/g)T^2}\,,\label{res}
\ee
with a  dimensionless function $\bar\eta(\bar B)$. We provide a full numerical result for this function in the massless $N_F=2$ QCD in section \ref{sec3}. In the limit $eB\to 0$, it reduces to the previously known shear viscosity without magnetic field in leading log order \cite{Arnold:2000dr}.

In this regime, the magnetic field appears only in the advective term in the effective Boltzmann kinetic theory, and its effects to the QCD collision term is of higher order and sub-leading compared to the usual
leading order collision term $\cal C$ which is already of order ${\cal C}\sim g^4\log(1/g) T$. This is because the dominant energy-momentum carriers responsible for the shear viscosity are hard particles of momentum $p\sim T$, and possible corrections from the magnetic field to the dispersion relation of hard particles are suppressed by $\sqrt{eB}/T\sim g^2$. It is by the same reason that the leading order collision term is obtained by using free dispersion relations of external hard particles without thermal corrections. 
Moreover, the screening mass (that is, the Debye mass) of order $m_D\sim gT$ that regulates IR divergences of t-channel collision terms  between these hard particles should also be the one without the magnetic field corrections. This is because the screening mass is provided by ``hard thermal 1-loop", that is, the screening is provided by hard particles themselves. Any correction to the dispersion relation of these hard particles is of order $\sqrt{eB}/T\sim g^2$, and therefore the correction to $m_D$ from magnetic field is naturally of order $m_D\times \sqrt{eB}/T\sim m_D g^2$. Since the net collision term is already of ${\cal C}\sim g^4 \log(1/g) T$, the correction to $\cal C$ from the correction to the screening mass is of higher order than this, which is not relevant in our computation.

Any remaining effect of $eB$ can only appear in the possible on-shell singularity in quark propagators with a long-lived intermediate fermion interacting with the background $eB$ field, that may have an IR enhancement. But, this is precisely what the advective term of the Boltzmann equation captures. The effect of background $eB$ on the long-lived charged quasi-particles is described by the Lorentz force in the advective term,
\be
{\partial f\over\partial t}+\hat{\bm p}\cdot{\partial f\over \partial\bm x}+\dot{\bm p}\cdot{\partial f\over\partial \bm p}={\cal C}[f]\,,\quad \dot{\bm p}=\pm q_F\hat{\bm p}\times (e\bm B)\,,
\ee
where $q_F=({2\over 3},-{1\over 3})$ for $(u,d)$-quarks, and $\pm$ refers to quark and anti-quark.

\begin{figure}[t]
	\centering
	\includegraphics[width=10cm]{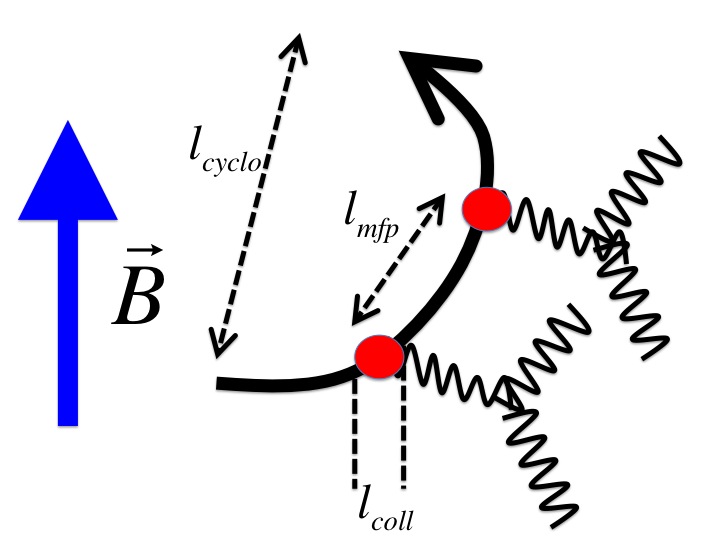}
		\caption{The meaning of the three length scales, $l_{coll}$, $l_{mfp}$ and $l_{cyclo}$.\label{fig0}}
\end{figure}
Another way of thinking about it is in terms of time scales. The space-time duration of each collision, $l_{coll}$, satisfies $T^{-1}\ll l_{coll}\ll (g T)^{-1}$, due to the fact that the log-enhancement in the t-channel 2-to-2 processes arises from the momentum exchanges lying between $gT$ and $T$. This time scale is much shorter than the time scale of cyclotron orbits induced by magnetic field which is $l_{cyclo}\sim p/(eB)\sim (g^4\log(1/g)T)^{-1}$, so that
each collisions look localized during any significant cyclotron motion, and are not affected by the magnetic field. This is why ${\cal C}$ in the Boltzmann equation remains the same. However, the inverse rate of large-angle collisions, or equivalently the mean-free distance {\it between} each collisions is $l_{mfp}\sim {\cal C}^{-1}\sim (g^4\log(1/g)T)^{-1}$, which is comparable to $l_{cyclo}$. This means that the cyclotron motions and the QCD collisions are equally important in the transport dynamics of hard particles. Therefore, one has to solve the above Boltzmann equation keeping both the Lorentz force term and the collision term $\cal C$, in order to capture the interplay between cyclotron motions and the QCD collisions. The dimensionless variable $\bar B$ in (\ref{res}) corresponds to the relative strength of the effect of cyclotron motions compared to the QCD collisions: $\bar B\sim l_{mfp}/l_{cyclo}$.

To compute the shear viscosity, we consider the kinetic theory of quarks/anti-quarks and gluons, whose distribution functions $(f^q,f^{\bar q},f^g)$ satisfy the Boltzmann equation with leading log collision term,
\bear
\partial_t f^q+\hat{\bm p}\cdot\partial_{\bm x}f^q+q_Fe(\hat{\bm p}\times\bm B)\cdot\partial_{\bm p}f^q&=&{\cal C}^q\,,\nonumber\\
\partial_t f^{\bar q}+\hat{\bm p}\cdot\partial_{\bm x}f^{\bar q}-q_Fe(\hat{\bm p}\times\bm B)\cdot\partial_{\bm p}f^{\bar q}&=&{\cal C}^{\bar q}\,,\nonumber\\
\partial_t f^g+\hat{\bm p}\cdot\partial_{\bm x}f^g&=&{\cal C}^g\,.
\eear
The equilibrium distribution with an arbitrary longitudinal velocity $u^\mu_\parallel=\gamma(1,v_z,\bm 0_\perp)$, $\gamma\equiv(1-v_z^2)^{-{1\over 2}}$ satisfies the detailed balance (that is, ${\cal C}=0$), and is a solution of the Boltzmann equation since $(\hat{\bm p}\times\bm B)\cdot\partial_{\bm p}f_{eq}=0$,
\be
f^{q,\bar q,g}_{eq}(\bm p,u^\mu_\parallel)=1/(e^{-{1\over T} p_\mu u^\mu_\parallel}\pm 1)\,,\quad p^\mu=(|\bm p|,\bm p)\,.
\ee
When we have a non-zero gradient of $u^\mu_\parallel$ or $v_z$ along the transverse space, $\partial_\perp v_z\neq 0$, the collision term which is local in space-time still vanishes with the above distribution, but the spatial gradient term in the advective term on the left no longer vanishes.
Considering the local rest frame of a point $x=0$ in space-time where the velocity is expanded in first order gradient as $v_z=(\partial_\perp v_z)x_\perp+{\cal O}(x_\perp^2)$, we have
\be
\hat{\bm p}\cdot\partial_{\bm x}f_{eq}(\bm p,u^\mu_\parallel(x))\Big|_{x=0}=(\hat{\bm p}_\perp\cdot\partial_\perp v_z) \beta p_z n_{F/B}(\bm p)(1\mp n_{F/B}(\bm p))\,,\quad \beta=1/T\,,
\ee
where $n_{F/B}$ is the Fermi-Dirac/Bose-Einstein distribution function
\be
n_{F/B}({\bm p})=1/(e^{\beta|\bm p|}\pm 1)\,.
\ee
This gradient term acts as a source for the disturbance of the distribution function away from the local equilibrium given by $f_{eq}(\bm p,u^\mu_\parallel(x))$.
The solution of the Boltzmann equation in linear order in $(\partial_\perp v_z)$ at the point $x=0$ is then
\be
f=f_{eq}(\bm p,u^\mu_\parallel(x))+\delta f(\bm p)\,,
\ee
where $\delta f$ satisfies the linearized Boltzmann equation
\bear
(\hat{\bm p}_\perp\cdot\partial_\perp v_z) \beta p_z n_{F}(\bm p)(1- n_{F}(\bm p))+q_Fe(\hat{\bm p}_\perp\times\bm B)\cdot\partial_{\bm p}\delta f^q&=&{\cal C}^q[\delta f^q,\delta f^{\bar q},\delta f^g]\,,\nonumber\\
(\hat{\bm p}_\perp\cdot\partial_\perp v_z) \beta p_z n_{F}(\bm p)(1- n_{F}(\bm p))-q_Fe(\hat{\bm p}_\perp\times\bm B)\cdot\partial_{\bm p}\delta f^{\bar q}&=&{\cal C}^{\bar q}[\delta f^q,\delta f^{\bar q},\delta f^g]\,,\nonumber\\
(\hat{\bm p}_\perp\cdot\partial_\perp v_z) \beta p_z n_{B}(\bm p)(1+ n_{B}(\bm p))&=&{\cal C}^g[\delta f^q,\delta f^{\bar q},\delta f^g]\,,\label{linear}
\eear
where we assume that a stationary state $\partial_t=0$ is achieved in the time scale much longer than the expected relaxation time $1/\tau_R\sim g^4\log(1/g)T$ that defines the boundary of the low energy hydrodynamics.

Once $\delta f$ is found from the above equations, the energy momentum tensor of our interest is computed as
\be
T^{\perp z}=\int {d^3\bm p\over (2\pi)^3}{\bm p_\perp p_z\over E_{\bm p}}\left(\nu_q\sum_F(\delta f^q+\delta f^{\bar q})+\nu_g \delta f^g\right)\,,\label{tzperp}
\ee
where $\nu_q=2N_c=2d_q$ and $\nu_g=2(N_c^2-1)=2d_A$ are the number of states of quarks/anti-quarks or gluons with a given momentum $\bm p$. ($d_q$ and $d_A$ are the dimensions of color representation). Comparing the result with (\ref{shear}), we obtain the shear viscosity $\eta$.

The leading log QCD collision term has been well known in literature starting from Refs.\cite{Baym:1990uj,Heiselberg:1994px} culminating in the full determination in Ref.\cite{Arnold:2000dr}. It was nicely re-derived and summarized in Ref.\cite{Hong:2010at} and we will follow the notations of Ref.\cite{Hong:2010at}.
There are two types of contributions to the leading log collision term: I) t-channel soft gluon exchanges and II) t-channel soft fermion exchanges. The type I processes do not change the particle species and give rise to diffusions in momentum space, whereas the type II processes convert fermions into gluons and vice versa.
Writing $\delta f$ as
\be
\delta f^a=n_{F/B}(1\mp n_{F/B})\chi^a\,,\quad a=q,\bar q, g\,,
\ee
the linearized collision term is obtained by
\be
{\cal C}^a[\delta f(\bm p)]=-{(2\pi)^3\over \nu_a}{\delta {\cal I}[\chi]\over\delta\chi^a(\bm p)}\,,\label{col}
\ee
where ${\cal I}={\cal I}^I+{\cal I}^{II}$ with
\bear
{\cal I}^I&=&{Tm_D^2g^2\log(1/g)\over 16\pi}\sum_a C_a\nu_a \int_{\bm p} n_a({\bm p})(1\mp n_a(\bm p))\left({\partial \chi^a(\bm p)\over\partial \bm p^i}\right)^2\nonumber\\
&-&{g^4\log(1/g)\over 16\pi d_A}\left(\sum_a C_a\nu_a \int _{\bm p} n_a(\bm p)(1\mp n_a(\bm p))\left(\hat{\bm p}\cdot{\partial\chi^a(\bm p)\over\partial\bm p}\right)\right)^2\nonumber\\
&-&{g^4\log(1/g)\over 16\pi d_A}\left(\sum_a C_a\nu_a \int _{\bm p} n_a(\bm p)(1\mp n_a(\bm p)){\partial\chi^a(\bm p)\over\partial\bm p^i}\right)^2\,,\label{I1}\\
{\cal I}^{II}&=& \gamma \sum_{a=q,\bar q} \nu_a \int_{\bm p}{1\over |\bm p|}{n_F({\bm p})(1+n_B({\bm p}))}\left(\chi^a(\bm p)-\chi^g(\bm p)\right)^2\nonumber\\&+&
{16\gamma\over T^2}\sum_{a=q} \nu_a \int_{\bm p}{1\over |\bm p|}n_F(\bm p)(1+n_B(\bm p))\left(\chi^a(\bm p)-\chi^g(\bm p)\right)\int_{\bm k}{1\over |\bm k|}n_F(\bm k)(1+n_B(\bm k))\left(\chi^{\bar a}(\bm k)-\chi^g(\bm k)\right)\nonumber\\&-&{8\gamma\over T^2}\sum_{a=q,\bar q}\nu_a \left(\int_{\bm p}{1\over |\bm p|}n_F(\bm p)(1+n_B(\bm p))\left(\chi^a(\bm p)-\chi^g(\bm p)\right)\right)^2\,,\label{I2}
\eear
where $C_a$ is the quadratic Casimir of the species $a$ and 
$\int_{\bm p}\equiv \int {d^3{\bm p}\over (2\pi)^3}$ and
\be
m_D^2={g^2 T^2\over 3}\left(N_c+{N_F\over 2}\right)\,,\quad \gamma={C_F^2T^2 g^4\log(1/g)\over 64\pi}\,.
\ee

Since the collision term is rotationally invariant, an inspection of the linearized Boltzmann equation (\ref{linear}) dictates the following form of the solution,
\bear
\chi^q(\bm p)&=&\left(\bm p_\perp\cdot \partial_{\perp} v_z\right)p_z\,\chi_+(|\bm p|)+\left(\bm p_\perp\times \partial_\perp v_z\right)p_z\,\chi_-(|\bm p|)\,,\nonumber\\
\chi^{\bar q}(\bm p)&=&\left(\bm p_\perp\cdot \partial_{\perp} v_z\right)p_z\, \chi_+(|\bm p|)-\left(\bm p_\perp\times \partial_\perp v_z\right)p_z \,\chi_-(|\bm p|)\,,\nonumber\\
\chi^g(\bm p)&=&\left(\bm p_\perp\cdot \partial_{\perp} v_z\right)p_z\,\chi_G(|\bm p|)\,,\label{ansatz}
\eear
where 
\be
\left(\bm p_\perp\cdot \partial_{\perp} v_z\right)\equiv {\bm p}_\perp^i\cdot \partial_{\bm x_\perp^i}v_z\,,\quad 
\left(\bm p_\perp\times \partial_{\perp} v_z\right)\equiv \epsilon^{ij}{\bm p}_\perp^i \partial_{\bm x_\perp^j}v_z\,,\quad i,j=x,y\,,
\ee
and the functions $\chi_\pm, \chi_G$ depend only on $p\equiv |\bm p|$. 
This is the most general form consistent with the isometry of the collision term and the residual $SO(2)_\perp$ rotational symmetry in the advective term in the presence of a background magnetic field.
Note that the second term in the first two equations involving $\chi_-$ comes from the Lorentz force term in (\ref{linear}) due to the magnetic field, and takes opposite sign between quark and anti-quark. When $B=0$, $\chi_-$ vanishes.

Inserting this form of the solution into (\ref{linear}) and working out the collision term (\ref{col}) explicitly,
one obtains a coupled set of second order differential equations for $\chi_\pm(p)$ and $\chi_G(p)$.
It is easy to see that only the first lines in (\ref{I1}) and (\ref{I2}) contribute to the final collision term, while the other terms vanish
for our solutions in (\ref{ansatz}). After some amount of algebra, we get the differential equations ($'\equiv {\partial\over\partial p}$ and $n_{F/B}\equiv n_{F/B}(p)$),
\bear
&&{1\over 8\pi}{Tm_D^2 g^2\log(1/g)}C_F\left([n_F(1-n_F)]'\left(2\chi_+(p)+p\chi_+'(p)\right)+n_F(1-n_F)\left(6\chi_+'(p)+p\chi_+''(p)\right)\right)\nonumber\\
&=&\beta n_F(1-n_F)+q_FeBn_F(1-n_F)\chi_-(p)+2\gamma n_F(1+n_B)\left(\chi_+(p)-\chi_G(p)\right)\,,\\
&&{1\over 8\pi}{Tm_D^2 g^2\log(1/g)}C_F\left([n_F(1-n_F)]'\left(2\chi_-(p)+p\chi_-'(p)\right)+n_F(1-n_F)\left(6\chi_-'(p)+p\chi_-''(p)\right)\right)\nonumber\\
&=&-q_FeBn_F(1-n_F)\chi_+(p)+2\gamma n_F(1+n_B)\chi_-(p)\,,\\
&&{1\over 8\pi}{Tm_D^2 g^2\log(1/g)}C_A\left([n_B(1+n_B)]'\left(2\chi_G(p)+p\chi_G'(p)\right)+n_B(1+n_B)\left(6\chi_G'(p)+p\chi_G''(p)\right)\right)\nonumber\\
&=&\beta n_B(1+n_B)+2\gamma {2N_c\over (N_c^2-1)} n_F(1+n_B)\sum_F \left(\chi_G(p)-\chi_+(p)\right)\,.
\eear

Once the solution is found, inserting (\ref{ansatz}) into (\ref{tzperp}) and performing angular integration using
\be
\int {d^3\bm p\over (2\pi)^3} { \bm p_\perp^i {\bm p}_\perp^j p_z^2\over E_{\bm p}}={\delta^{ij}\over 2}
\int {d^3\bm p\over (2\pi)^3} { \bm p_\perp^2p_z^2\over E_{\bm p}}
={\delta^{ij}\over 2(2\pi)^2}\int_{-1}^{+1}dx\, x^2(1-x^2) \int_0^\infty dp \,p^5={\delta^{ij}\over 30\pi^2} \int_0^\infty dp \,p^5\,,
\ee
we have the shear viscosity
\be
\eta=-{1\over 30\pi^2}\int_0^\infty dp \,p^5\left(2(N_c^2-1)n_B(1+n_B)\chi_G(p)+\sum_F 4N_c n_F(1-n_F)\chi_+(p)\right)\,.
\ee

\section{Results for shear viscosity in magnetic field \label{sec3}}

It is more convenient to formulate the equations in the previous section in terms of dimensionless variable $\bar p\equiv p/T$. In our numerical analysis in this work, we 
take a simplifying approximation of considering two identical flavors with a same charge $\bar q_F$,
and replace $\sum_F\to N_F=2$. We expect that this will not affect the major features of our result presented below, and leave a full consideration of the case of different charges in the future. At least we know that the sign of the charge should not matter for shear viscosity, since the effects from magnetic field appear in combination of $q_F^2(eB)^2$ due to charge-conjugation symmetry.
Defining the dimensionless magnetic field strength introduced in the introduction
\be
\bar B\equiv {|\bar q_F eB|\over g^4\log(1/g)T^2}\,,
\ee
with the similar definitions for dimensionless quantities
\be
\bar m_D^2\equiv m_D^2/T^2\,,\quad \bar\gamma\equiv\gamma/(T^2g^4\log(1/g))\,,
\ee
and the dimensionless functions 
\be
\bar\chi_\pm\equiv T^3 g^4\log(1/g) \chi_\pm\,,\quad\bar\chi_G\equiv T^3 g^4\log(1/g) \chi_G\,,
\ee
we have the dimensionless differential equations ($'\equiv {\partial\over\partial\bar p}$)
\bear
&&{1\over 8\pi}{\bar m_D^2 }C_F\left([n_F(1-n_F)]'\left(2\bar\chi_++\bar p\bar\chi_+'\right)+n_F(1-n_F)\left(6\bar\chi_+'+\bar p\bar\chi_+''\right)\right)\nonumber\\
&=&n_F(1-n_F)+\bar Bn_F(1-n_F)\bar \chi_-+2\bar\gamma n_F(1+n_B)\left(\bar\chi_+-\bar\chi_G\right)\,,\label{de1}\\
&&{1\over 8\pi}{\bar m_D^2 }C_F\left([n_F(1-n_F)]'\left(2\bar\chi_-+\bar p\bar\chi_-'\right)+n_F(1-n_F)\left(6\bar\chi_-'+\bar p\bar\chi_-''\right)\right)\nonumber\\
&=&-\bar Bn_F(1-n_F)\bar\chi_++2\bar\gamma n_F(1+n_B)\bar\chi_-\,,\label{de2}\\
&&{1\over 8\pi}{\bar m_D^2 }C_A\left([n_B(1+n_B)]'\left(2\bar\chi_G+\bar p\bar\chi_G'\right)+n_B(1+n_B)\left(6\bar\chi_G'+\bar p\bar\chi_G''\right)\right)\nonumber\\
&=&n_B(1+n_B)+2\bar\gamma {2N_cN_F\over (N_c^2-1)} n_F(1+n_B) \left(\bar\chi_G-\bar\chi_+\right)\,,\label{de3}
\eear
in terms of which the shear viscosity is written as
\be
\eta=\bar\eta(\bar B){T^3\over g^4\log(1/g)}\,,
\ee
with the dimensionless function
\be
\bar\eta(\bar B)=-{1\over 30\pi^2}\int_0^\infty{d\bar p}\, \bar p^5\left(2(N_c^2-1)n_B(1+n_B)\bar\chi_G(\bar p)+4N_c N_F n_F(1-n_F)\bar\chi_+(\bar p)\right)\,.\label{shearfinal}
\ee

In principle we can solve numerically the above inhomogeneous second order differential equations imposing regular boundary conditions at $\bar p=0$ and $\bar p=\infty$ that uniquely determine the solution for given $\bar B$. However we find that it is not practically easy to do this using the shooting method since we have three functions and we need to scan three dimensional parameter space of initial conditions.
For the case of $B=0$, we have $\chi_-=0$ and we can barely manage to find the solution scanning the reduced two dimensional space of initial conditions. We get in this way $\bar\eta=86.46$ for $B=0$ and $N_F=2$. For $N_F=0$ where both $\chi_+$ and $\chi_-$ vanish, we get $\bar\eta=27.12$. These results agree very well with the previous results by Arnold-Moore-Yaffe \cite{Arnold:2000dr}, and give us confidence that our differential equations in the above do not contain trivial mistakes.

It is possible to solve the above equations in the limiting case of $\bar B\to\infty$.
It can be shown without difficulty that in this limit, 
\be
\bar\chi_+\sim {\cal O}(1/\bar B^2)\,,\quad \bar\chi_-\sim{\cal O}(1/\bar B)\,,\quad \bar\chi_G\sim{\cal O}(1)\,,
\label{largeB}\ee
so that it is enough to solve the last equation (\ref{de3}) for $\bar\chi_G$ neglecting $\bar\chi_+$ to get the limiting value of $\bar\eta(\infty)$. This can easily be done by the shooting method, and we obtain 
\be
\bar\eta(\bar B)=18.87+{\cal O}(1/\bar B^2)\,,\quad\bar B\to\infty\,.
\ee

Instead of solving the differential equations by the shooting method, we follow Ref.\cite{Arnold:2000dr} to 
find approximate solutions within a finite dimensional functional space formed by a well-chosen set of basis functions. We first show that our original problem of solving the differential equations and computing the integral (\ref{shearfinal}) for the shear viscosity is equivalent to a variational problem of a quadratic action.
Defining the ``Lagrangians'' ${\cal I}_+$, ${\cal I}_-$, ${\cal I}_G$ and ${\cal I}_{\rm mix}$ by
\bear
{\cal I}_+&=& -{4N_c N_F\over 30\pi^2}\left({1\over 8\pi}\bar m_D^2C_Fn_F(1-n_F) \left(\bar p^2([\bar p^2 \bar\chi_+]')^2+6\bar p^4 \bar\chi_+^2\right)+2\bar\gamma\bar p^5n_F(1+n_B)\bar\chi_+^2\right)\,,\nonumber\\
{\cal I}_-&=& +{4N_c N_F\over 30\pi^2}\left({1\over 8\pi}\bar m_D^2C_F n_F(1-n_F)\left(\bar p^2([\bar p^2 \bar\chi_-]')^2+6\bar p^4 \bar\chi_-^2\right)+2\bar\gamma\bar p^5n_F(1+n_B)\bar\chi_-^2\right)\,,\nonumber\\
{\cal I}_G&=& -{2(N_c^2-1)\over 30\pi^2}\left({1\over 8\pi}\bar m_D^2C_A n_B(1+n_B)\left(\bar p^2([\bar p^2 \bar\chi_G]')^2+6\bar p^4 \bar\chi_G^2\right)+2\bar\gamma{2N_cN_F\over (N_c^2-1)}\bar p^5n_F(1+n_B)\bar\chi_G^2\right)\,,\nonumber\\
{\cal I}_{\rm mix}&=&-{1\over 15\pi^2}4N_c N_F\left( -2\bar\gamma \bar p^5 n_F(1+n_B)\bar\chi_+\bar\chi_G+\bar B \bar p^5 n_F(1-n_F)\bar\chi_+\bar\chi_-\right)\,,\label{lagrangian}
\eear  
and the source by
\be
{\cal S}=-{1\over 15\pi^2}\left(4N_cN_F \bar p^5n_F(1-n_F)\bar\chi_++2(N_c^2-1)\bar p^5n_B(1+n_B)\bar\chi_G\right)\,,
\ee 
we consider the action functional
\be
\bar\eta[\bar\chi_+,\bar\chi_-,\bar\chi_G]\equiv \int_0^\infty\,d\bar p\,\,\left({\cal I}_++{\cal I}_-+{\cal I}_G+{\cal I}_{\rm mix}+{\cal S}\right)\,,
\ee
which is at most quadratic in $\bar\chi$'s.
It is then straightforward to show that the equations of motions from the above action coincide with the differential equations in (\ref{de1}), (\ref{de2}), and (\ref{de3}), so that the solution of the differential equations corresponds to the extrema of the action functional. Moreover the value of the action evaluated at the extrema is equal to the dimensionless function $\bar\eta(\bar B)$ in (\ref{shearfinal}), that is
\be
\bar\eta(\bar B)=\bar\eta[\bar\chi_+,\bar\chi_-,\bar\chi_G]\Big|_{\rm solution}\,.
\ee

\begin{figure}[t]
	\centering
	\includegraphics[width=10cm]{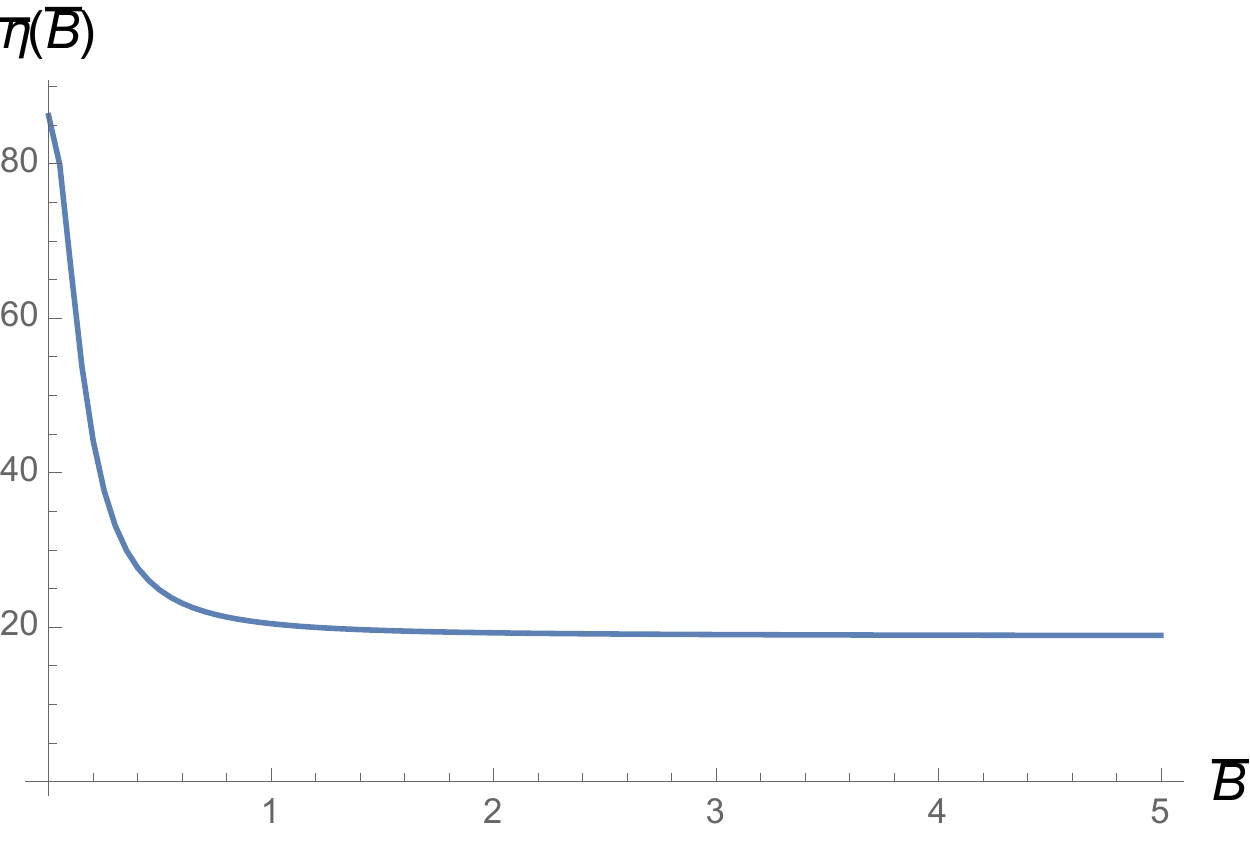}
		\caption{The numerical result for $\bar\eta(\bar B)$ with $N_F=2$.\label{fig1}}
\end{figure}
We perform this variational analysis in a finite dimensional functional space spanned by a set of basis functions $\phi^{(m)}(\bar p)$ ($m=1,\ldots, N$): writing $\bar\chi_\pm$ and $\bar\chi_G$ as
\be
\bar\chi_\pm(\bar p)=\sum_{m}a^{(m)}_\pm\phi^{(m)}(\bar p)\,,\quad \bar\chi_G(\bar p)=\sum_m b^{(m)}\phi^{(m)}(\bar p)\,,
\ee
with variational coefficients $(a^{(m)}_\pm,b^{(m)})\equiv A$, we evaluate the action to find
\be
\bar\eta[A]=-{1\over 2}A^T {\cal I}A+A^T {\cal S}\,,
\ee
with a $(3N)\times (3N)$ symmetric matrix ${\cal I}$ and a $(3N)$-dimensional column vector ${\cal S}$. Solving for the extremum by $A={\cal I}^{-1}{\cal S}$ and evaluating the action at the extremum, we get
the approximate result for the shear viscosity
\be
\bar\eta(\bar B)={1\over 2}{\cal S}^T{\cal I}^{-1}{\cal S}\,.
\ee
Following Ref.\cite{Arnold:2000dr} we choose the basis functions
\be
\phi^{(m)}(\bar p)={\bar p^{m-1}\over (1+\bar p)^{N-1}}\,,\quad m=1,\ldots,N\,,
\ee
with $N$ up to 10 which was claimed to produce a good precision in the case of $B=0$.
In our case with $B\neq 0$, we expect the numerical results to be less accurate, since the action functional 
$\bar\eta[\bar\chi_+,\bar\chi_-,\bar\chi_G]$ with $B\neq 0$ is not bounded above and the extremum doesn't correspond to a global maximum of $\bar\eta$: note that ${\cal I}_-$ in (\ref{lagrangian}) is positive definite, while ${\cal I}_+$ and ${\cal I}_G$ is negative definite. Our numerical result is estimated to be trustable within $\pm 1\%$ level.
Figure \ref{fig1} shows our numerical result of $\bar\eta(\bar B)$ for $N_F=2$. Starting from $\bar\eta(0)\approx 86.46$, it rapidly decreases up to $\bar B=1$, beyond which
$\bar\eta(\bar B)$ decreases slowly to the limiting value $\bar\eta(\infty)=18.87$. 

\section{Discussion}

We emphasize that our result is limited to the case where the strength of magnetic field is weak compared to the temperature, with a parametric dependence $eB=\bar B g^4\log(1/g)T^2$ where $\bar B$ is a dimensionless number. We also assume that the electromagnetism is non-dynamical and external, which means that the magnetohydrodynamics (MHD) regime is never realized. With these assumptions, our leading log value of shear viscosity takes the following scaling form $\eta=\bar\eta(\bar B)T^3/(g^4\log(1/g))$, where $\bar\eta(\bar B)$ depends only on the ratio $\bar B=eB/(g^4\log(1/g)T^2)$. 
Our result could give a useful insight on the effects of magnetic field on the shear viscosity of quark-gluon plasma in late stages of the heavy-ion collisions where the magnetic field becomes weak compared to temperature scale. However, a more realistic study of the transport coefficients in the MHD regime with five shear and bulk viscosities classified in Refs.\cite{Huang:2011dc,Hernandez:2017mch} still remains demanding.

A peculiar feature in our result shown in Figure \ref{fig1} is a steep fall of the value of $\bar\eta(\bar B)$ within a narrow region of $\bar B$ near zero. This can be understood as follows. 
First of all, the limiting {\it finite} value at $\bar B\to\infty$ comes solely from the contribution of gluons which are neutral and have no cyclotron orbits, whereas the quarks and anti-quarks do not contribute to the transport in this limit because their cyclotron orbits become very small compared to the collisional mean free path. This can be explicitly seen in (\ref{largeB}). Note that the transport of gluons still feel the presence of quarks and anti-quarks via the collisions with the background quarks and anti-quarks (that is reflected in the value of the Debye mass $m_D^2$ in the collision terms depending on the number of flavors), and this is why the shear viscosity in this limit is not equal to the one in $N_F=0$ case. Indeed, in a pure gluon case of $N_F=0$, we checked that if one simply replaces the Debye mass in the $N_F=0$ case with the one in $N_F=2$ case, one obtains precisely our limiting value of 18.87 for the shear viscosity.

Therefore, the question is why the quark and anti-quark contributions drop quickly.
We first recall from our discussion around Figure \ref{fig0} that their contributions would drop significantly when the size of the cyclotron orbit $l_{cyclo}$ becomes comparable to the collisional mean free path $l_{mfp}$.
Denoting the characteristic size of the collision term as ${\cal C}\sim C\,g^4 \log(1/g) T$ with a dimensionless number $C$, we have $l_{mfp}\sim {\cal C}^{-1}\sim C^{-1}/(g^4 \log(1/g)T)$. On the other hand, the cyclotron orbit size is $l_{cyclo}\sim p/(eB)\sim \bar B^{-1} /(g^4\log(1/g)T)$ with our dimensionless parameter $\bar B$. This means that their ratio is
\be
 l_{mfp}/l_{cyclo}\sim \bar B/C\,,
 \ee
and therefore we expect that $\bar\eta(\bar B)$ will drop significantly around $\bar B\sim C$.
Now the shear viscosity at $\bar B=0$ is simply proportional to $l_{mfp}\sim {\cal C}^{-1}\sim C^{-1}/(g^4\log(1/g)T)$ (times kinematic factors): in other words, the numerical value of $\bar\eta(0)$ should be roughly given by $C^{-1}$. It is just a feature of explicit QCD collision terms that $C$ is a numerically small number: for example, it should be about $1/80$ to give the value $\bar\eta(0)=86.46$ for $N_F=2$. This would mean that an order 1 value of $\bar B$ is already in the regime where $l_{mfp}/l_{cyclo}\gg 1$ and the quark contribution should be negligible. In summary, the numerical smallness of the QCD collision term in unit of $g^4\log(1/g)T$ is responsible for both a large value of $\bar\eta(0)$ {\it and} the narrowness of the $\bar B$ dependence.
This inverse correlation between $\bar\eta(0)$ and the width in $\bar B$ is universal, and can also be found, for example, in the Drude picture of transport in magnetic field.

As a future direction, it seems possible to go one step further and compute the shear viscosity in soft magnetic field in complete leading order, using the variational approach as in Ref.\cite{Arnold:2003zc}.
It will also be interesting to compute the bulk viscosities in magnetic field appearing in the constitutive relation (\ref{consti}). See Ref.\cite{Hattori:2017qih} for recent development.

\vskip 1cm \centerline{\large \bf Acknowledgment} \vskip 0.5cm
This work is supported by the U.S. Department of Energy, Office of Science, Office of Nuclear Physics, with the grant No. DE-SC0018209 and within the framework of the Beam Energy Scan Theory (BEST) Topical Collaboration.

\section*{Appendix}

In this Appendix, we show that the most general tensor structures for the first order viscous part of energy-momentum tensor in a magnetic field classified in Refs.\cite{Huang:2009ue,Huang:2011dc} precisely reduce to our expression in Eq.(\ref{consti}), when the fluid velocity is restricted to be only longitudinal along the magnetic field direction, say $\hat z$, and the plasma is neutral. As a by-product, we will be able to relate the five shear and two bulk viscosoties in Ref.\cite{Huang:2011dc} with our three viscosities in Eq.(\ref{consti}).

Let us start from the most general viscous part of the energy-momentum tensor classified, e.g. in Eq.(39) in Ref.\cite{Huang:2011dc} (the signs of a few terms are reversed due to our different metric convention, see Eq.(B7) in Ref.\cite{Finazzo:2016mhm} for a direct comparison with the same metric convention with ours):
\bear
\tau^{\mu\nu}&=&-2\eta_0\left(w^{\mu\nu}-{1\over 3}\Delta^{\mu\nu}\theta\right)-\eta_1\left(\Delta^{\mu\nu}-{3\over 2}\Xi^{\mu\nu}\right)\left(\theta-{3\over 2}\phi\right)\nonumber\\&+&2\eta_2\left(b^\mu\Xi^{\nu\alpha}b^\beta+b^\nu\Xi^{\mu\alpha}b^\beta\right)w_{\alpha\beta} +2\eta_3\left(\Xi^{\mu\alpha}b^{\nu\beta}+\Xi^{\nu\alpha}
b^{\mu\beta}\right)w_{\alpha\beta}\nonumber\\
&-& 2\eta_4(b^{\mu\alpha}b^\nu b^\beta+b^{\nu\alpha}b^\mu b^\beta)w_{\alpha\beta}-{3\over 2}\zeta_\perp \Xi^{\mu\nu}\phi-3\zeta_\parallel b^\mu b^\nu\varphi\,,\label{classif}
\eear
where $\Delta^{\mu\nu}=u^\mu u^\nu+g^{\mu\nu}$ is the projection operator orthogonal to the velocity $u^\mu$, $w^{\mu\nu}={1\over 2}\Delta^{\mu\alpha}\Delta^{\nu\beta}(\partial_\alpha u_\beta+\partial_\beta u_\alpha)={1\over 2}(\Delta^\mu u^\nu+\Delta^\nu u^\mu)$ is the projected shear tensor ($\Delta^\mu=\Delta^{\mu\alpha}\partial_\alpha$), $\theta=\partial_\mu u^\mu$ is the expansion, $b^\mu=B^\mu/|B|$ is the unit vector ($b_\mu b^\mu=1$) proportional to the magnetic field $B^\mu={1\over 2}\epsilon^{\mu\nu\alpha\beta}u_\nu F_{\alpha\beta}$, $\Xi^{\mu\nu}=\Delta^{\mu\nu}-b^\mu b^\nu$ is the transverse projection operator orthogonal to both $u^\mu$ and $b^\mu$ (note that $b^\mu$ is already orthogonal to $u^\mu$), $b^{\mu\nu}=\epsilon^{\mu\nu\alpha\beta}u_\alpha b_\beta$ is the unit normalized field strength tensor (that is, $b^{\mu\nu}=F^{\mu\nu}/|B|$), $\phi=\Xi^{\mu\nu}w_{\alpha\beta}$ and $\varphi=b^\mu b^\nu w_{\alpha\beta}$ are the transverse and longitudinal expansions respectively.
Our convention is $g^{\mu\nu}={\rm diag}(-1,+1,+1,+1)$ and $\epsilon^{txyz}=+1$.

Now, let's fix the direction of the magnetic field to be $\vec B=B\hat z$ (or $F_{xy}=B$), and assume that the fluid velocity has only the longitudinal components, $u^\mu=u^\mu_\parallel=\gamma(1,v_z,0,0)$, where the first two components are for $(t,z)$ directions and the last two for $(x,y)$ ($\gamma=(1-v_z^2)^{-{1\over 2}}$ as usual, so that $u_\parallel^\mu u_{\parallel\mu}=-1$). With this general $u^\mu_\parallel$ vector, we can still keep the 1+1 dimensional Lorentz covariance in the above expression, whereas the transverse boosts that will change the background magnetic field into electric field are not useful, and we don't intend to keep covariance for them. Let us introduce some of our notations: we decompose the metric into its longitudinal and transverse components by $g^{\mu\nu}=g^{\mu\nu}_\parallel+g^{\mu\nu}_\perp$ where $g^{\mu\nu}_\parallel={\rm diag}(-1,+1,0,0)$ and $g^{\mu\nu}_\perp={\rm diag}(0,0,+1,+1)$, then it is easy to see that $g^{\mu\nu}_\perp$ can be identified with the projection operator to the transverse $(x,y)$ space, $\Delta^{\mu\nu}_\perp=g^{\mu\nu}_\perp$. We also introduce the longitudinal projection operator that is perpendicular to the velocity and the transverse space, $\Delta_\parallel^{\mu\nu}=u^\mu_\parallel u^\nu_\parallel+g_\parallel^{\mu\nu}$. Then the full projection operator is the sum of the above two: $\Delta^{\mu\nu}=u^\mu_\parallel u^\nu_\parallel+g^{\mu\nu}=\Delta_\parallel^{\mu\nu}+\Delta_\perp^{\mu\nu}$. We define the purely transverse gradient as $\partial_\perp^\mu=\Delta^{\mu\nu}_\perp\partial_\nu=g^{\mu\nu}_\perp \partial_\nu$. We introduce the 1+1 dimensional antisymmetric tensor $\epsilon_\parallel^{\mu\nu}$ where it is non-zero only when $(\mu,\nu)$ is along $(t,z)$ with the convention $\epsilon_\parallel^{tz}=+1$. Similarly, we introduce the transverse antisymmetric tensor in $(x,y)$ directions $\epsilon_\perp^{\mu\nu}$ with $\epsilon_\perp^{xy}=+1$.

With these at hand, we now describe several simplifications happening when the velocity is purely longitudinal.
All the results below can most easily be checked by working out explicitly in the local rest frame where $u^\mu_\parallel=(1,0,0,0)$, and since these equations are covariant under 1+1 dimensional Lorentz transformations, they must be true in the lab frame as well. However, it would be also useful to give some general argument for these simplifications as we do in the following.
The $b^\mu$ is a longitudinal 1+1 dimensional vector as the magnetic field and the velocity are also longitudinal, and at the same time it is perpendicular to $u^\mu_\parallel$. In 1+1 dimensions, the unique unit vector (up to a sign) that is perpendicular to $u^\mu_\parallel$ is $\epsilon_\parallel^{\mu\nu}u_{\parallel \nu}$ (this is simply because there are not enough number of dimensions in 1+1 dimensions), so $b^\mu$ should be equal to this up to a sign: $b^\mu={\rm sgn}(B)\epsilon_\parallel^{\mu\nu} u_{\parallel\nu}$. This also means that the longitudinal projection operator $\Delta^{\mu\nu}_\parallel$ perpendicular to $u_\parallel^\mu$ can be constructed from $b^\mu$, that is,  $\Delta^{\mu\nu}_\parallel=u^\mu_\parallel u^\nu_\parallel+g_\parallel^{\mu\nu}=b^\mu b^\nu$. Then, the projection $\Xi^{\mu\nu}$ becomes $\Xi^{\mu\nu}=\Delta^{\mu\nu}-b^\mu b^\nu=\Delta^{\mu\nu}_\parallel+\Delta_\perp^{\mu\nu}-b^\mu b^\nu=\Delta^{\mu\nu}_\perp=g_\perp^{\mu\nu}$, that is a purely transverse projection operator. Also from $b^{\mu\nu}=F^{\mu\nu}/|B|$, it can be written as $b^{\mu\nu}={\rm sgn}(B)\epsilon^{\mu\nu}_\perp$. As a final and most non-trivial simplification, we show that the following is true:
\be
\Delta^{\mu\alpha}_\parallel \Delta^{\nu\beta}_\parallel \partial_\alpha u_{\parallel\beta}=\Delta^{\mu\nu}_\parallel \theta\,.
\ee
This can be shown in the local rest frame, where $\mu=\nu=\alpha=\beta=z$ for a non-zero value in the left-hand side, and $\theta=\partial_\mu u^\mu_\parallel=\partial_z u^z_\parallel$ since $u_{\parallel\mu}\partial_\nu u_\parallel^\mu=0$ for all $\nu$ (from $u_{\parallel\mu}u^\mu_\parallel=-1$) means that $\partial_\nu u^t_\parallel=0$ in the local rest frame. Note also that $\theta=\partial_\mu u^\mu_\parallel=\partial_{\parallel\mu}u^\mu_\parallel$ where $\partial^\mu_\parallel=\Delta^{\mu\nu}_\parallel\partial_\nu$. The above equation which is covariant in 1+1 dimensions should then be true in any frame.

With these simplifications, it is straightforward to derive the following reduction of the tensors appearing in (\ref{classif}),
\bear
w^{\mu\nu}-{1\over 3}\Delta^{\mu\nu}\theta&=&{1\over 2}(\partial_\perp^\mu u_\parallel^\nu+\partial_\perp^\nu u_\parallel^\mu)+{2\over 3} \Delta^{\mu\nu}_\parallel \theta-{1\over 3}g^{\mu\nu}_\perp\theta\,,\nonumber\\
\Delta^{\mu\nu}-{3\over 2}\Xi^{\mu\nu}&=&\Delta^{\mu\nu}_\parallel-{1\over 2}g^{\mu\nu}_\perp\,,\nonumber\\
\left(b^\mu\Xi^{\nu\alpha}b^\beta+b^\nu\Xi^{\mu\alpha}b^\beta\right)w_{\alpha\beta}&=&{1\over 2}(\partial_\perp^\mu u_\parallel^\nu+\partial_\perp^\nu u_\parallel^\mu)\,,\nonumber\\
\left(\Xi^{\mu\alpha}b^{\nu\beta}+\Xi^{\nu\alpha}
b^{\mu\beta}\right)w_{\alpha\beta}&=&0\,,\nonumber\\
(b^{\mu\alpha}b^\nu b^\beta+b^{\nu\alpha}b^\mu b^\beta)w_{\alpha\beta}&=&{1\over 2}{\rm sgn}(B)\left(\epsilon^{\mu\alpha}_\perp\partial_{\perp\alpha}u_{\parallel}^\nu+\epsilon^{\nu\alpha}_\perp\partial_{\perp\alpha}u_{\parallel}^\mu\right)\,,\nonumber\\
\Xi^{\mu\nu}\phi&=&0\,,\nonumber\\  b^\mu b^\nu \varphi&=&
\Delta^{\mu\nu}_\parallel\theta\,.
\eear
Using these, the viscous part of the energy-momentum tensor (\ref{classif}) greatly reduces to
\bear
\tau^{\mu\nu}&=&-\eta(\partial_\perp^\mu u_\parallel^\nu+\partial_\perp^\nu u_\parallel^\mu)-(\zeta \Delta^{\mu\nu}_\parallel+\zeta' g^{\mu\nu}_\perp)\theta\nonumber\\
&+& \eta_4 \,\,{\rm sgn}(B)\left(\epsilon^{\mu\alpha}_\perp\partial_{\perp\alpha}u_{\parallel}^\nu+\epsilon^{\nu\alpha}_\perp\partial_{\perp\alpha}u_{\parallel}^\mu\right)\,,
\eear
where the first line agrees with precisely what we have in our Eq.(\ref{consti}) with the identification 
\be
\eta=\eta_0-\eta_2\,,\quad \zeta=3\zeta_\parallel+{4\over 3}\eta_0+\eta_1\,,\quad \zeta'=-{2\over 3}\eta_0-{1\over 2}\eta_1\,.
\ee
It is interesting that the transverse bulk viscosity $\zeta'$ in response to longitudinal expansion can be negative (that is, the transverse pressure can increase when the longitudinal expansion is positive) when $2\eta_0+\eta_1>0$, but we don't have a good understanding on this.
The second line in the above is also interesting: since it is charge-conjugation (C) odd due to the fact that the magnetic field is C-odd, it vanishes in a neutral plasma (the same is true for $\eta_3$ from the fact that both $b^\mu$ and $b^{\mu\nu}$ are C-odd). To understand this clearly, the second line term for a non-relativistic velocity gives a contribution to $T^{zi}$ ($i=x,y$) as
\be
T^{zi}=\eta_4\,\,{\rm sgn}(B)\epsilon^{ij}_\perp\partial_{\perp j}v_z\,,
\ee
which is a momentum transfer {\it perpendicular} to the gradient of velocity in the $(x,y)$ plane.
It is not difficult to understand this in terms of Lorentz force acting on microscopic charged particles that are transported in the direction of velocity gradient, and since this Lorentz force is opposite between quarks and anti-quarks, their contributions to this momentum transport coefficient cancel with each other in a neutral plasma.

 \vfil

\end{document}